\documentclass [12pt]{article}
\usepackage{graphicx,amsmath}

\begin{document}
\title{Comparative Dielectric Behavior of PbFe$_{1 / 2}$Ta$_{1 /
2}$O$_{3}$ and NaNbO$_{3}$:Gd Relaxor-Like Crystals}

\author{I. P. Raevski$^{1,2}$, S. A. Prosandeev$^{1,3}$, U. Waghmare
$^{3,4}$, \\V. V. Eremkin$^{2}$ , V. G. Smotrakov$^{2}$ , V. A.
Shuvaeva$^{2}$}

\maketitle

$^{1}$Physics Department, Rostov State University, 344090 Rostov
on Don, Russia

$^{2}$ Research Institute of Physics, Rostov State University,
344090 Rostov on Don, Russia

$^{3}$Ceramics Division, A317/223, 100 Bureau Drive, STOP 8520,
National Institute of Standards and Technology, Gaithersburg, MD

$^{4}$Theoretical Sciences Unit, J Nehru Centre for Advanced,
Scientific Research, Bangalore, India 560 064

\begin{abstract}
Experimental data obtained for PbFe$_{1 / 2}$Ta$_{1 / 2}$O$_{3 }$
(PFT) and NaNbO$_{3}$: Gd (NaNbGd) single crystals show a diffused
dielectric permittivity peak that is inherent to relaxor
ferroelectrics. However some deviations from the normal relaxor
properties were also observed and are under discussion. One of the
key features of the relaxors is the existence of the Burns
temperature, at which the polar regions appear. We found out that
PFT shows this feature but the NaNbGd dielectric behavior is
different. We analyse these properties within a phenomenological
theory and propose a microscopic model.
\end{abstract}

\section{Introduction}

Relaxor properties of heterogeneous ferroelectrics are still
intriguing although the main phenomenon, the diffuseness of the
temperature dielectric-permittivity peak, was discovered long ago
\cite{1}. It is now established that one of the important features
of relaxors is the polar-regions formation and these regions grow
with decreasing temperature and reach their maximal sizes at the
ferroelectric or glass-type phase transition \cite{2}.

New information, rather useful for understanding the relaxor
properties, was recently obtained from neutron scattering which
establishes the existence of a soft ferroelectric mode above the
Burns temperature $T_{d}$, and below the freezing temperature
$T_{g }(T_{d} > T_{g})$, \cite{3,4}. In the intermediate
temperature range, the ``waterfall phenomenon'' is seen \cite{4}
which reflects the absence of transverse optical mode below a
definite wave vector. Another new idea stemmed from the diffuse
scattering of neutrons in relaxors suggesting a uniform
displacement of the ions in polar regions along with optical
displacements \cite{5}. We use these findings to establish the
order parameter in the relaxor state.

The present study considers the dielectric behavior of PbFe$_{1 /
2}$Ta$_{1 / 2}$O$_{3 }$ (PFT), and NaNbO$_{3}$:Gd (NaNbGd)
crystals which have some common features (the diffuseness of the
temperature dielectric peak) as well as some differences. In PFT,
the extrapolated Curie-Weiss temperature, $T_{CW}$, is above the
temperature of dielectric permittivity maximum, $T_m $, hinting
that a ferroelectric phase transition would occur if the crystal
were homogeneous. The heterogeneities in the crystal prevent the
appearance of the ferroelectric phase and lead to changes in the
temperature dependence of dielectric permittivity characteristic
of the relaxor state below $T_{CW}$ \cite{1}. We connect these
changes with the appearance of a new order parameter corresponding
to a nanodomain structure and show an example of such a structure
within a model considering a heterogeneous material consisting of
slabs (for the sake of simplicity) having different elastic and
dielectric properties. In contrast with PFT, we find that $T_{CW
}$is lower than $T_m $ and another order parameter appears at $T_m
$ in NaNbGd.

\section{Experimental}

PbFe$_{1 / 2}$Ta$_{1 / 2}$O$_{3 }$(PFT), and NaNbO$_{3}$:Gd
(NaNbGd) crystals were grown by the flux method. The crystals
obtained had an isometric (edge dimensions 0.5 -2 mm) form with
the sides parallel to the {\{}100{\}} planes of the perovskite
prototype lattice. The details of the crystal preparation and
characterization have been described elsewhere \cite{6,7}.

The dielectric studies were carried out in the 10$^{3}$ -10$^{6 }$Hz range
in the course of both heating and cooling at the rate about 2 -3 K/min with
the aid of the R5083 and E7-12 capacitance bridges. Aquadag electrodes for
dielectric measurements were deposited on the opposite faces of the as-grown
crystals.

\section{PbFe$_{1 / 2}$Ta$_{1 / 2}$O$_{3}$}

At high temperatures PFT has the cubic perovskite structure in
which the B-perovskite-position is occupied by the approximately
randomly distributed Fe$^{3 + }$ and Ta$^{5 + }$ ions. This random
distribution influences the displacements of the Pb ions and can
result in the appearance of polar clusters whose interactions and
growths can trigger a phase transition into a glassy or ordered
phase. Fig. 1 shows the real part of the dielectric permittivity
of the PFT crystal measured at different frequencies in the
10$^{3}$ -10$^{6 }$Hz range. The strong frequency dispersion of
\textit{$\varepsilon $} observed at temperatures well above
$T_{d}$ is likely to be due to conductivity \cite{1}. In addition
to the known \textit{$\varepsilon $(T)} curves for PFT crystals
\cite{8} and ceramics \cite{1,9}, besides a diffused and
frequency-dependent \textit{$\varepsilon $(T)} maximum typical of
relaxors, an inflection is observed in the \textit{$\varepsilon
$(T)} curve at temperature $T_{ip}$, which is 30-40 K lower then
$T_{m}$. The frequency dispersion of \textit{$\varepsilon $} is
less pronounced at temperatures below this inflection than above
it. Such behavior is similar to that observed in PbSc$_{1 /
2}$Ta$_{1 / 2}$O$_{3 }$ and PbSc$_{1 / 2}$Nb$_{1 / 2}$O$_{3 }$
\cite{10} at the spontaneous transition from the relaxor to normal
ferroelectric state (see also \cite{11}), but for PFT, it can be
also connected with a glass-type phase transition which appears
due to percolation of polarization \cite{12,13}, or
antiferromagnetic phase transition \cite{1,8} (recent results on
Rietveld refinement of the neutron diffraction data \cite{14} does
not show any symmetry changes in PFT from room temperature down to
10 K when the superstructure due to the antiferromagnetic ordering
is taken into account).

The fit of the Curie-Weiss law to the high temperature side of the
\textit{$\varepsilon $(T)} curve measured at a high frequency
(Fig. 2) provides the extrapolated Curie-Weiss temperature $T_{CW}
\approx $ 310 K which is substantially higher than $T_m $. In the
same figure we show the difference between the experimental $1 /
\varepsilon $ curve and the Curie-Weiss fit. It is seen that this
difference has two portions where it behaves approximately
linearly with temperature. One of the portions intersects the
temperature axis at temperature $T_\eta $ which approximately
equals $T_{CW}$, while the other, seen at lower temperatures,
intersects the $T$ axis at a temperature which is close to the
value $T_{VF} \approx 246\,\,\mbox{K}$~ obtained from the
Vogel-Fulcher fit of the $T_m$~ dependence on frequency.

The fact that $T_\eta \approx T_{CW} $ hints to identify a new order
parameter appearing at this temperature with local polarization. It is
consistent with the idea that in PFT, at some temperature, $T_{d}$, there
appear polar regions ($T_{d}$ is higher than $T_{CW}$ because of the diffusness
of the phase transition that we will discuss in detail in Section 5). At the
same temperature the dielectric permittivity starts deviating from the
Curie-Weiss law.

It was rigorously shown in Ref. \cite{15} that, in the
heterogeneous media consisting of a ferroelectric slab and
dielectric layers (dead layers) the ferroelectric region gets
broken into 180$^{0}$ domains in order to decrease the
depolarization field (see Section 6 where we consider the rigorous
solution for similar inhomogeneous problem: two ferroelectric
slabs separated by dielectric interfacial layers). On this basis
we propose the appearance of such (possibly short-range) randomly
oriented domain structure in the relaxor state below $T_{d}$. We
will discuss the consistency of this idea with other experiments
in Section 7.

\section{NaNbO$_{3}$:Gd}

NaNbO$_{3}$ exhibits an antiferroelectric phase transition at
630$^{0}$ K \cite{1,16}. When being doped with Gd at small
concentrations the temperature of this phase transition decreases
and the width of the temperature hysteresis increases. We
discussed these trends in Ref. \cite{16} assuming the (1-x)
NaNbO$_{3 }$--(x)Gd$_{1 / 3}$NbO$_{3}$~ solid solution formation.
At $x \approx 0.12$ the thermal hysteresis abruptly disappears and
a diffused hysteresis-free dielectric peak remains at higher
concentrations (Fig. 3). We regarded this behavior in Ref.
\cite{16} to the appearance of a concentration phase transition
into the relaxor-like state. However there are features which are
different in comparison with ordinary relaxors.

First of all we find that that the $T_{CW}$ found from the
Curie-Weiss fit of the high-temperature side of the dielectric
permittivity is much lower than $T_{m}$ (Fig. 4). This makes
scenario described in the previous section not suitable for the
experimental data for NaNbGd and another scenario should be
developed. In the next Section we will consider a reason for the
diffusness of the phase transition. We will show that, in spite of
the different meaning of the order parameter $\eta $ for PFT and
NaNbGd, the diffuseness of the dielectric permittivity peak in
these two cases can have similar origin.

\section{Landau-type theory of the diffuseness of the phase
transitions in PFT and NaNbGd}

In Ref. \cite{16} we considered the strain and average square of
local polarization as concomitant order parameters to an
antiferroelectric order parameter. The addition of Gd results in
the appearance of some distorted polar regions. The order
parameter which is connected with these regions and responsible
for the deviation from the Curie-Weiss behavior can have different
meaning in different materials. For example, it can be the average
square of polarization or local deformation or, if one thinks
about the electronic subsystem, the degree of electron
localization. In the case of PFT this order parameter can
correspond to the magnitude of the local polarization in the
nanodomain structure. In NaNbGd it can be an antiferroelectric
order parameter. We use for this order parameter the notation
$\eta $, we assume that it is scalar or a tensor of the second
kind, which in the product with $P^{2}$ gives a scalar, and we
write Landau Free energy expansion:

\begin{equation}
\label{eq1}
\begin{array}{l}
 F = F_0 + \frac{1}{2}\left( {\alpha + qu} \right)P^2 + \frac{1}{4}\beta P^4
+ \frac{1}{6}\gamma P^6 + \frac{1}{2}\lambda P^2\eta + \\
 + \frac{1}{2}A\eta ^2 + \frac{1}{3}B\eta ^3 - ku\eta + cu^2 - EP - (\sigma
_0 + \sigma )u + ... \\
 \end{array}
\end{equation}

\noindent
where $\alpha = a(T - T_{CW} )$, $A = b(T - T_\eta )$. Here $T_\eta \ge
T_{CW} $; $u $is strain and $\sigma _0 $ is internal stress (caused by Gd in the
case of NaNbGd).

The equilibrium solutions for this Free energy are simple:

\begin{equation}
\label{eq2}
\begin{array}{l}
 1.\,P = 0,\,\eta = 0 \\
 2.\,P = 0,\,\eta = \dfrac{ - A + \sqrt {A^2 + 4kuB} }{2B} \\
 3.\,\eta = \dfrac{ - A + \sqrt {A^2 + 4B(ku + \lambda P^2)} }{2B},\\P^2 =
\dfrac{\alpha + qu + \lambda \eta \pm \sqrt {\left( {\alpha + qu +
\lambda \eta } \right)^2 - 4\beta \gamma } }{2\gamma } \\
 \end{array}
\end{equation}

Here $u$~ is the strain found from the equilibrium condition (it
is always finite because of the stress $\sigma _0 $; the
electrostriction contribution to the Hamiltonian
\textit{quP}$^{2}$ simply shifts the Curie-Weiss temperature and
changes the value of the nonlinearity constant $\beta )$. The
first phase is stable at high temperatures. Then, at lower
temperatures, the phase 2 becomes stable, with the order parameter
$\eta $, and, at even lower temperatures, the third phase, where
the ferrolectric order parameter (here we do not pay much
attention to the difference between the glass-type and
ferrolectric solutions as we are mostly interested in the relaxor
state) coexists with the order parameter $\eta $. The values of
$\eta $ and $P$ can be found from the solution of equations
(11.3).

In order to find the dielectric permittivity one can write the equilibrium
condition with respect to $P$:

\begin{equation}
\label{eq3}
\left( {\alpha + qu} \right)P + \beta P^3 + \gamma P^5 + \lambda P\eta = E
\end{equation}

By taking the derivative of this equation with respect to $E$ one
can find

\begin{equation}
\label{eq4}
\chi = \frac{1}{\varepsilon _0 }\frac{\partial P}{\partial E} =
\frac{1}{\varepsilon _0 }\frac{1}{\alpha + qu + 3\beta P^2 + 5\gamma P^4 +
\lambda \eta }
\end{equation}

For the phase (2) in (\ref{eq2}) $P = $0 and $\lambda \eta =
\lambda b\left( {\sqrt {x^2 + \delta } - x} \right) / 2B$ where $x
= (T - T_\eta )$, $\delta = 2ku / b$. In this case (\ref{eq4})
takes the form

\begin{equation}
\label{eq5}
\varepsilon = \frac{1}{\varepsilon _0 }\frac{1}{\alpha + qu + \lambda
b\left( {\sqrt {x^2 + \delta } - x} \right) / 2B}
\end{equation}

From experimental data we found $\varepsilon _0 \lambda \eta $
which is simply the difference between $1 / \varepsilon $ and the
high temperature Curie-Weiss dependence $\left( {1 / \varepsilon }
\right)_{CW} = \varepsilon _0 \left( {\alpha + qu} \right)$. A fit
of the expression $\varepsilon _0 \lambda \eta = \varepsilon _0
\lambda b\left( {\sqrt {x^2 + \delta } - x} \right) / 2B$ to the
experimentally obtained $1 / \varepsilon - \left( {1 / \varepsilon
} \right)_{CW} $ for NaNbGd is shown in Fig. 5 (Notice that the
same quality fit and the same results were obtained from the fit
of (\ref{eq5}) to the corresponding experimental curve). We find
that this fit obtained with $\varepsilon _0 a = 3.2861\cdot 10^{ -
6}\pm 3.78\cdot 10^{ - 9}$, $T_{CW} = 185.75\pm 1.84$ K,
$\varepsilon _0 \lambda b / 2B = 3.91\cdot 10^{ - 6}\pm 3\cdot
10^{ - 8}$ K$^{ - 1}$, $T_\eta = 308.4\pm 0.6$ K and $\delta =
1084.4\pm 54$ K$^{2}$ is rather good and explains the deviation of
the dielectric permittivty from the Curie-Weiss law by the
appearance of the new order parameter below $T_\eta $ and by the
existence of local stresses. We have obtained a similar good fit
for PFT (Fig. 5) at the values: $\varepsilon _0 \lambda b / 2B =
7.16\cdot 10^{ - 6}\pm 1.2\cdot 10^{ - 7}$ K$^{ - 1}$, $T_\eta =
T_{CW} = 295.6\pm 1.0$ K and $\delta = 1450.6\pm 80$ K$^{2}$. From
these data the appearance of the quadratic temperature
dependencies of $\varepsilon (T)$ at the maximal position becomes
clear: it appears due to the expansion of the square root in
expression (\ref{eq3}) with respect to $x$ for small $x$. Hence we
regard the diffuseness of the temperature maximum of the
dielectric permittivity in NaNbGd to local quenched stresses and
fields produced by the Gd impurities in the matrix of NaNbO$_{3}$
and local polar fields produced by the random distribution of Fe
and Ta at the B-sites in PFT.

\section{A model for structural inhomogeneity}

We propose a model for structural inhomogeneity based on the
recent finding by Bratkovskiy and Levanyuk \cite{15} who studied a
short-circuited ferroelectric slab with dead layers. Consider
short-circuited ferroelectric slabs of the width $f $(for the sake
of simplicity we consider two slabs) divided by dielectric
interfacial layers with the width $d_{0}$ at the boundaries of the
whole system (were the system has metalic contacts) and with the
width $d$ in the middle (see Fig. 6).

We show that the strain can be excluded from the equations by the ordinary
minimization procedure. Let us consider the term describing the
electrostriction effect together with the corresponding elastic
contribution: $H_{elast} (r) = c_{\alpha \beta \alpha \beta } u_{\alpha
\beta }^2 / 2 - q_{zz\alpha \beta } u_{\alpha \beta } P_z P_z $ where
$u_{\alpha \beta } (r)$ is a component of the strain and $P_z (r)$ is a
component of local polarization (at finite values of the local polarization
one may consider a piezoelectric effect instead of the electrostriction one
taking into account that $\left( {P_s + \delta P} \right)^2 \approx P_s^2 +
2P_s \delta P)$. Minimizing this Hamiltonian with respect to the strain
component one immediately obtains: $u_{\alpha \beta } = q_{zz\alpha \beta }
P_z P_z / c_{\alpha \beta \alpha \beta } $ (at finite local polarizations
$u_{\alpha \beta } = d_{z\alpha \beta } P_z / c_{\alpha \beta \alpha \beta }
$ where $d $is a piezoelectric coefficient depending on the direction of the
local polarization). It implies that the strain can be excluded from the
total Hamiltonian for problems with zero or constant stress. Then, we look
for a lowest energy solution with possible spatial alternation of
polarization for ferroelectrics which are improper ferroelastics. Detailed
solution is worked out in Appendix A.

The macroscopic field energy (see Appendix, expression (A8)) is represented
in the form

\begin{equation}
\label{eq6}
f_M = F_M / LS = f_{M0} + \textstyle{1 \over 2}A\sigma _0^2 - E^\ast \sigma
_0
\end{equation}

\noindent where $\sigma _0 $ is the average charge on the
ferroelectric-dielectric boundary; the coefficients are given in
Appendix A. The effective field $E^{\ast }$ differs from the
average field $E = U / \varepsilon _t L$ by the factor
$2\varepsilon _t^2 f / \varepsilon _{2z} L$ where $U$ is applied
voltage, $\varepsilon _t $ is the total dielectric permittivity
obtained treating the system as a series of the capacitors,
$\varepsilon _{2z} $ is the dielectric permittivity of the
ferroelectric slab. This factor is small only when $f $(thickness
of the ferroelectric clab) is very small. This fact implies that
the effective acting field $E^{\ast }$ conjugated with $\sigma _0
$ for the thick ferroelectric layers and thin dielectric
interfacial layers is largely enhanced with respect to the average
field that makes it possible to change the polarization by very
small \textit{dc} fields.

On the basis of these results we predict the appearance of the
alternating domains in the ferroelectric slabs separated by
dielectric layers in lamellar structures. Perhaps some evidence
for this was found in the KTaO$_{3}$/KNbO$_{3}$ superstructure
\cite{17} although this has not been understood well yet.

The equilibrium value of the macroscopic charge at the boundaries $\sigma _0
$ corresponds to the minimum of the macroscopic Free energy: $\sigma _0 =
U\varepsilon _8 / 8\pi L_g $ ($\varepsilon _g $ is the dielectric
permittivity of the capacitors corresponding only to the dielectric
interfacial layers, $L_{g}$ is the total width of the dielectric layers along
thefield). Correspondingly, the derivative of this charge with respect to
field gives

\begin{equation}
\label{eq7}
\frac{d\sigma _0 }{dE} = \frac{L\varepsilon _g }{8\pi L_g }
\end{equation}

It is seen that this derivative is large if the total relative
length of the dielectric interfacial layers $L_{g}/L$~is small.
This result is consistent with that obtained in \cite{15} for one
ferroelectric slab having dead layers. There can be some
variations of the dielectric region's widths but only their total
relative width is important. We point out that this contribution
to the dielectric permittivity arises only from the domain walls
movements due to the voltage $U $which results in change in total
polarization, hence on $\sigma _0 $.

We now relate the stripe-like solution to experimentally
observable quantities. For a given solution of local polarization,
we obtain local strains through the minimization of energy with
respect to local strain \cite{34} as described at the beginning of
this section. The local strains are then used to obtain the
acoustic-like (collective motions of atoms in the unit cell)
atomic displacements $d_\alpha \left( r \right)$ through the
relation: $u_{\alpha \beta } = \frac{1}{2}\left( {\frac{\partial
d_\alpha }{\partial r_\beta } + \frac{\partial d_\beta }{\partial
r_\alpha }} \right)$. For simplicity, we restrict here to analysis
in 2 dimensions.

In addition to the solution with polarization in the domain
perpendicular to the plane of the strip (the transverse case T)
described in this section, we consider a longitudinal (L) case
where the polarization is in the plane of stripe and along the
direction perpendicular to the domain wall. Note that the
polarization in the adjacent domains is opposite in sign, giving
180$^{0}$-type domain walls separating them. The polarization is
along (1-1) and (-11) directions for the transverse case and along
(\ref{eq2}) and (-1-1) directions in the longitudinal case, and
the strip itself runs along (\ref{eq2}) direction. In Figure 7, we
show contourplots of atomic displacements, local strains and local
polarization for the two cases. It is clear that the atomic
displacements in the transverse case are almost constant at the
boundaries between polar domains and the embedding dielectric. In
contrast, we find large accumulated local strains at the domain
walls for the longitudinal case, making it costly in energy.

We use atomic displacements $d_\alpha $(acoustic modes) and
polarization $P_\alpha $(optical modes) in Eq. (2) of \cite{13} to
obtain inelastic scattering intensity plots, displayed in Fig. 8.
The scattering plots for both T and L resemble those shown in
Figure 4 of Ref. \cite{5}. The atomic displacements $d_\alpha $
corresponding to the acoustic modes (inhomogeneous strain) in our
analysis corresponds to the uniform shifts that Ref. \cite{5} uses
to model their data. Our results that the patterns are rather
similar for both T and L cases implies that more than one kind of
atomic displacements or structures could be used to model
experimental results and that the stripe-like solutions for the
inhomogeneity is consistent with the experimental observation.

\section{Discussion {\&} Summary}

We have presented new experimental data on perovskite crystals
which show diffuse temperature dielectric anomalies. We presented
a Landau-type theory of this diffuseness by considering local
stresses produced by inhomogeneities. In the case of PFT the
principal finding is that the extrapolated Curie temperature
$T_{CW}$ is larger than $T_{m}$ and coincides with the temperature
$T_\eta $. Following the recently proposed idea describing
alternating domains in a heterogeneous media consisted of a
ferroelectric thin film having thin dead layers or an epitaxial
film grown on a substrate \cite{15,19} we propose that similar
(but, most likely, short-range and random) structures appear in
relaxors below $T_{d}$. We provided a rigorous solution for an
extended heterogeneous media, two ferroelectric slabs divided by
dielectric interfacial layers, and obtained that this lamellar
structure also must have alternating domains in the ferroelectric
slabs. We showed that the contribution of the domain wall
movements into the dielectric permittivity in such a system is
governed by the relative ratio of the total width of the
dielectric layers and does not depend on the distribution of the
widths over the layers. We also computed strains, displacements
and inelastic scattering intensity of such a domain structure
which are in line with recent experiments \cite{3,4,5,20}. The
assumption that there are only two main kinds of the regions,
ferroelectric and dielectric, is consistent with the experimental
finding showing the existence of two different transverse optical
modes corresponding to a two-band structure of heterogeneous
binary compounds \cite{3}.

The alternating domain structure proposed is consistent with many
experiments. Indeed, for instance, the ``waterfall phenomenon''
\cite{4,20} could be related in this case to (nanoscale) domains,
the domain size determines the critical wave vector below which
the waterfall phenomenon is observed. The coupling between local
strain and polarization in alternating domains leads to atomic
displacements consistent with the inelastic scattering results
\cite{5} and the observation of strong interaction between
acoustic and optical modes \cite{3,20}. A strong coupling between
the ferroelectric fluctuations and domain wall fluctuations leads
to strong damping of ferroelectric fluctuations below the critical
wave vector.

We have shown that in the model of the domain structure considered
the contribution of the domain walls movements into dielectric
response is especially large in the case when the relative width
of the dielectric layers in the direction of the field is small
and it is consistent with the recent theoretical result obtained
for a ferroelectric thin film having dead layers \cite{15}. The
universal relaxation recently observed in PMN even at temperatures
larger than $T_{m}$ \cite{21} can now be understood on the basis
of the irreversible domain wall movements \cite{22} (the movements
of the domain walls until there exists an electric field without
further restoration after switching the field off). The Debye-type
relaxation of domain walls is expected in the cases when there is
a strong restoring field and this is also consistent with the
existence of a strong dispersion at $T_{m}$ \cite{12,22,23}. The
dielectric permittivity can now be found on the basis of the
theory considering the phonons coupled to relaxators
\cite{24,25,26,27}. This coupling in addition to the domains'
freezing can explain the experimentally evidenced deviation from
the Arrhenius law \cite{25,26,27}. Some contribution to the
dielectric permittivity from the Maxwell-Wagner mechanism is also
possible.

Formation of the proposed alternating domain structure is not
hampered by the depolarization field which is unfavorable for the
formation of lone polar regions in a dielectric media \cite{28}.
The cooperative appearance of the alternating polar regions make
them energetically more favorable than the lone separated polar
regions \cite{15}. We should stress that the lone polar regions
(for example in chemical clusters) can also exist and contribute
to the dielectric permittivity in the presence of free electrons
that compensate the depolarization fields \cite{29}; the change of
polarization in this case is caused by both the local ionic dipole
reorientations \cite{30} and by the compensating charge movements.

A picture emerging from the data analysis is the following. At
$T_{d}$ there appear precursors of the alternating nanodomain
structure with a short range alternation of the local
polarization. The maximal possible wave vector of this domain
structure just corresponds to the critical wave vector seen in the
``waterfall'' phenomenon \cite{4,20}. The dielectric permittivity
of relaxors is high due to the large ferroelectric fluctuations
and due to domain walls movements (see also
\cite{22,23,28,31,32}). At some temperature, known as the freezing
temperature, the nanodomains become very large most probably due
to the polarization percolation \cite{12} and are frozen
thereafter or a ferroelectric phase transition can take place.
This scenario seems to be valid in the case of PMN and other
classic relaxor materials as well.

In the case of NaNbGd $T_{CW}$ lies below $T_{m}$, and the presence of the
diffused temperature dielectric permittivity maximum, in our opinion, is due
to the local stresses and local fields produced by the Gd impurities. The
large difference between $T_\eta $ and $T_{CW} $ in this case hints that the
new order parameter appearing at $T_\eta $ has nothing to do with
ferroelectric domain structure, also resulting in much smaller permittivity.

\section{Acknowledgements}

The authors appreciate discussions with Gehring, Shirane, Bokov,
and Bratkovskiy. The study has been partially supported by RFBR
grants 01-02-16029 and 01-03-33119.

\section*{Appendix A}

The Poisson equation inside each of the slabs can be written in
the form

\begin{equation}
\begin{array}{l}
 {\varphi }''_{1x} + {\varphi }''_{1y} + {\varphi }''_{1z} =
0;\,\,\,{\varphi }''_{5x} + {\varphi }''_{5y} + {\varphi }''_{5z} =
0;\,\,\,{\varphi }''_{3x} + {\varphi }''_{3y} + {\varphi }''_{3z} = 0; \\
 \varepsilon _{2x} {\varphi }''_{2x} + \varepsilon _{2y} {\varphi }''_{2y} +
\varepsilon _{2z} {\varphi }''_{2z} = 0;\varepsilon _{2x} {\varphi }''_{4x}
+ \varepsilon _{2y} {\varphi }''_{4y} + \varepsilon _{2z} {\varphi }''_{4z}
= 0. \\
\end{array}
\label{A1}
\end{equation}

Here $\varepsilon _1 ,\varepsilon _3 $ are the dielectric permittivities in
the dielectric layer at the boundary of the system and in the middle
respectively; $\varepsilon _{2z} ,\varepsilon _{2x} ,\varepsilon _{2y} $ are
the dielectric permittivities in the ferroelectric slab in the corresponding
directions; $\varphi _i $ is the potential in the i-th slab, ${\varphi
}'_{i\alpha } $ and ${\varphi }''_{i\alpha } $ are the first and the second
derivatives with respect to $\alpha $. The boundary conditions between the
slabs at the voltage $U$ are:

\begin{equation}
\begin{array}{l}
 \varepsilon _{2z} {\varphi }'_{2z} ( - f / 2) - \varepsilon _1 {\varphi
}'_{1z} (d_0 ) = - 4\pi \sigma ;\,\,\,\varepsilon _3 {\varphi }'_{3z} ( - d
/ 2) - \varepsilon _{2z} {\varphi }'_{2z} (f / 2) = 4\pi \sigma \\
 \varepsilon _{2z} {\varphi }'_{4z} ( - f / 2) - \varepsilon _3 {\varphi
}'_{33z} (d / 2) = - 4\pi \sigma ;\,\,\,\varepsilon _1 {\varphi }'_{5z} (d_0
) - \varepsilon _{2z} {\varphi }'_{41z} (f / 2) = 4\pi \sigma \\
 \varphi _1 (0) = - U / 2;\,\,\,\varphi _1 (d_0 ) = \varphi _2 ( - f /
2);\,\,\,\varphi _2 (f / 2) = \varphi _3 ( - d / 2);\,\,\, \\
 \varphi _3 (d / 2) = \varphi _4 ( - f / 2);\,\,\,\varphi _4 (f / 2) =
\varphi _5 (d_0 );\,\,\,\varphi _5 (0) = U / 2 \\
\end{array}
\label{A2}
\end{equation}

\noindent where the general solutions are given by \cite{33}

\begin{equation}
\begin{array}{l}
\varphi_1(z,q_x,q_y) = C_{11} \dfrac{\sinh k_1 z}{\sinh k_1 d_0 }
+ C_{12} \dfrac{\cosh (k_1 z)}{\cosh (k_1 d_0 )}\,\,0 < z < d_0 \\
 \varphi_2 (z_2 ,q_x ,q_y ) = C_{21} \dfrac{\sinh (k_2 z_2 )}{\sinh (k_2 f /
2)} + C_{22} \dfrac{\cosh (k_2 z_2 )}{\cosh (k_2 f / 2)},\,\,\, -
f / 2 < z_2 < f / 2 \\
 \varphi_3 (z_3 ,q_x ,q_y ) = C_{31} \dfrac{\sinh (k_3 z_3 )}{\sinh (k_3 d /
2)},\,\,\, - d / 2 < z_3 < d / 2 \\
 \varphi_4 (z_4 ,q_x ,q_y ) = C_{41} \dfrac{\sinh (k_2 z_4 )}{\sinh (k_2 f /
2)} + C_{42} \dfrac{\cosh (k_2 z_4 )}{\cosh (k_2 f / 2)},\,\,\, -
f / 2 < z_4 < f / 2 \\
 \varphi_5 (z_5 ,q_x ,q_y ) = C_{51} \dfrac{\sinh k_1 z_5 }{\sinh k_1 d_0 }
+ C_{52} \dfrac{\cosh (k_1 z_5 )}{\cosh (k_1 d_0 )},\,\,\,0 < z_5
< d_0 \\
\end{array}
\label{A3}
\end{equation}

Here $z_2 = z + d_0 + f / 2$, $z_3 = z_2 + f / 2 + d / 2$. The
interfacial charge can be represented as a wave corresponding to
alternating domains in the $x$~ and $y$~ directions placed in the
checkboard (it is close to the honeycomb structure) or stripe-type
style:

\begin{equation}
 \begin{array}{l}
 \sigma (x,y) = \sum\limits_{q_x ,q_y } {\sigma (q_x ,q_y )e^{q_x x + q_y
y}} \\
 \sigma = \dfrac{2P_s }{iq_x T_x }\left( {1 - e^{iq_x a_{x1} }} \right)\delta
_{q_y 0} ,\,\,q_x \ne 0,\,\,\,\,\,stripes \\
 \sigma = - \dfrac{4P_s }{q_x T_x q_y T_y }\left( {1 - e^{iq_x a_{x1} }}
\right)\left( {1 - e^{iq_y a_{1y} }} \right),\,\,\,q_x^2 + q_y^2 \ne
0;\,\,\,checkboard \\
 \end{array}
 \label{A4}
\end{equation}

Here $P_{s}$ is the local polarization magnitude inside domains; at zero wave
vector $\sigma = \sigma _0 $ where $\sigma _0 $ can be found from the
equilibrium condition which we will derive below. The lengths of the domains
in the $x $and $y $directions ($a_{\alpha 1} $ and $a_{\alpha 2} $ where $\alpha =
x,y$ for polarization up and down, $T_\alpha = a_{\alpha 1} + a_{\alpha 2}
)$ depend on the electric field magnitude.

Above we considered only the case when the polarization in the ferroelectric
slabs is parallel, as, in the antiparallel case, the energy of the system is
larger. The solution of the above equations in this case has the following
symmetry conditions: $C_{42} = - C_{22} ;\, \quad \,C_{41} = C_{21} $; $C_{52} =
- C_{12} ;\,\,\,\,C_{51} = - C_{11} $. From the equality of the potentials
at the interfaces (see conditions (A2)) one obtains: $C_{31} = - C_{21} -
C_{22} ; \quad C_{11} = - C_{21} + C_{22} - C_{12} $; $C_{12} = 0$ at $q_x^2 +
q_y^2 \ne 0$ and $C_{12} = - U / 2$ at $q_x^2 + q_y^2 = 0$. From the Poisson
equations in (A2) we have $C_{21} F_1 - C_{22} F_2 - C_{11} G_0 = - 4\pi
\sigma $ and $C_{31} G_1 - C_{21} F_1 - C_{22} F_2 = 4\pi \sigma $, where
$G_0 = \varepsilon _1 k_1 \coth \left( {k_1 d_0 } \right)$, $\,G_1 =
\varepsilon _3 k_3 \coth \left( {k_3 d / 2} \right)$, $F_1 = \varepsilon
_{2z} k_2 \coth \left( {k_2 f / 2} \right)$, $F_2 = \varepsilon _{2z} k_2
\tanh \left( {k_2 f / 2} \right)$. Excluding $C_{31 }$and $C_{11}$ one
has:$(F_1 + G_0 )C_{21} - (F_2 + G_0 )C_{22} = - 4\pi \sigma - C_{12} G_0 $,
$ - (F_1 + G_1 )C_{21} - (F_2 + G_1 )C_{22} = 4\pi \sigma $. Under these
conditions we obtained the following solution (only one coefficient will be
necessary for the Free energy calculation) for $q_x^2 + q_y^2 \ne 0$

\begin{equation}
C_{21} = - \frac{4\pi \sigma \left( {2F_2 + G_0 + G_1 }
\right)}{(F_1 + G_0 )(F_2 + G_1 ) + (F_1 + G_1 )(F_2 + G_0
)}
\label{A5}
\end{equation}

\noindent
and for $q_x^2 + q_y^2 \to 0$:

\begin{equation}
C_{21} = \left( { - 4\pi \sigma \frac{L_g }{\varepsilon _g } + U}
\right)\frac{\varepsilon _t f}{2\varepsilon _{2z} L}
\label{A6}
\end{equation}

Here $L / \varepsilon _t = L_g / \varepsilon _g + L_f / \varepsilon _{2z} $,
$L_g / \varepsilon _g = d / \varepsilon _3 + 2d_0 / \varepsilon _1 $, $L_f =
2f$.

The $k_{1}$, $k_{2}$ and $k_{3}$ can be found from the Poisson equations:
$k_1^2 = k_3^2 = q_x^2 + q_y^2 , \quad \varepsilon _{2z} k_2^2 = \varepsilon
_{2x} q_x^2 + \varepsilon _{2y} q_y^2 $. Finally one obtains the Free energy
in the form:

\begin{equation}
F = F_0 - 2S\sum\limits_{q_x ,q_y } {C_{21} \sigma (q_x ,q_y )} -
\frac{S\varepsilon _t U^2}{8\pi L}
\label{A7}
\end{equation}

\noindent where $S$~ is the surface area covered by the domains.
The first term includes the Landau expansion with respect to
polarization, domain wall energy and the energy connected with the
appearance of polarization under the bias field; the second term
describes the electrostatic energy of the boundaries; and the last
term describes the capacitor energy due the bias field.

From (\ref{A7}) the macroscopic field energy (the terms
corresponding to $q_x^2 + q_y^2 = 0$) can be represented in the
form

\begin{equation}
f_M = F_M / LS = f_{M0} + \textstyle{1 \over 2}A\sigma _0^2 -
E^\ast \sigma _0
\label{A8}
\end{equation}

\noindent
where $f_{M0} = \varepsilon _t U^2 / 8\pi L^2$, $A = 8\pi L^{ - 1}\left(
{\varepsilon _{2z} / L_f + \varepsilon _g / L_g } \right)^{ - 1}$, $E^\ast =
\varepsilon _t L_f U / \varepsilon _{2z} L^2$.

The alternating domain period can be now found from the equilibrium
condition for the alternating field contribution to the Free energy

\begin{equation}
F_{alt} = \frac{Nf\Delta SP_s^2 }{a} + \sum\limits_{q_x^2 + q_y^2
\ne 0} {\frac{8\pi S\sigma ^2\left( {2F_2 + G_0 + G_1 }
\right)}{\left( {F_1 + G_0 } \right)\left( {F_2 + G_1 } \right) +
\left( {F_1 + G_1 } \right)\left( {F_2 + G_0 } \right)}}
\label{A9}
\end{equation}

\noindent
where $N = $4 in the case of the checkboard order ($a = a_{x1} = a_{x2} = a_{y1} =
a_{y2} )$ and $N = $2 in the case of the striped domains ($a = a_{x1} = a_{x2}
;\,\,\,a_{y1} = a_{y2} = \infty )$; $\Delta $ is the characteristic domain
wall width.

Especially simple solution can be obtained in the case when the dielectric
and ferroelectric layers are thick in comparison with the domain width
[15,34] and $\varepsilon _{2x} = \varepsilon _{2y} $:

\begin{equation}
a = \sqrt {\frac{Nf\Delta \left( {\varepsilon _{2g} + \varepsilon
_1 } \right)\left( {\varepsilon _{2g} + \varepsilon _3 }
\right)}{64\xi \left( {2\varepsilon _{2g} + \varepsilon _1 +
\varepsilon _3 } \right)}} \label{A10}
\end{equation}

\noindent
where $\varepsilon _{2g} = \sqrt {\varepsilon _{2z} \varepsilon _x } $; $\xi
$ does not depend on $a $and for the checkboard order we have:

\begin{equation}
\frac{\pi ^4\xi }{4} = \sum\limits_{i = 0}^\infty {\sum\limits_{j
= 0}^\infty {\frac{1}{\left( {2i + 1} \right)^2\left( {2j + 1}
\right)^2\sqrt {\left( {2i + 1} \right)^2 + \left( {2j + 1}
\right)^2} }} } \approx 0.813
\label{A11}
\end{equation}

In the case of the striped domains \cite{34}

\begin{equation}
 \frac{\pi ^2\xi }{2} = \sum\limits_{i = 0}^\infty
{\frac{1}{\left( {2i + 1} \right)^3}} \approx 1.0518
\label{A12}
\end{equation}

The domain width increases with dielectric permittivity and the thickness of
the ferroelectric slab.

In the case of thin dielectric interfacial layers and
comparatively thick ferroelectric slabs in the stripe-type order
and at $\varepsilon _1 = \varepsilon _3 = \sqrt {\varepsilon _{2z}
\varepsilon _{2x} } $, $d/$2$ = d_{0}$~ the problem can be reduced
to the case considered in Ref. \cite{15}, and the final result is

\begin{equation}
a = 0.95d\exp \left( {0.4\frac{a_K^2 }{d^2}} \right)\label{A13}
\end{equation}

\noindent where $a_{K}$ is the Kittel domain size parameter which
coincides with (\ref{A10}). This expression shows that the domains
extremely rapidly grow when the relative size of the dielectric
interfacial layers in the direction of field becomes thinner.

\section*{Figure captions}

Figure 1. \textit{$\varepsilon $'(T)} dependencies for the PFT
crystal, measured at different frequencies.

Figure 2. Temperature dependencies of \textit{$\varepsilon $'} (1) and 1/$\varepsilon $' (2)
measured at 10$^{6 }$Hz for the PFT crystal. Curve 3 shows the difference
between the experimental $1 / \varepsilon $ dependence and the Curie-Weiss
fit of $1 / \varepsilon $ . Straight solid lines are guides to the eye.

Figure 3. Temperature dependencies of \textit{$\varepsilon $'}
measured at 10$^{3}$, 10$^{4}$,10$^{5}$, 10$^{6 }$Hz. for
0.88NaNbO$_{3}$-0.12Gd$_{1 / 3 }$NbO$_{3}$ crystal.

Figure 4. Temperature dependencies of \textit{$\varepsilon $}
and \textit{1/$\varepsilon $} measured at 10$^{5 }$Hz for
0.88NaNbO$_{3}$-0.12Gd$_{1 / 3 }$NbO$_{3}$ crystal and the difference
between the experimental $1 / \varepsilon $ dependence and the Curie-Weiss
fit of $1 / \varepsilon $. Straight solid lines are guides to the eye.

Figure 5. The fit of the model expression to the experimental data.

Figure 6. The model lamellar structure consisted of ferroelectric and
dielectric slabs.

Figure 7.Contourplots of (a) atomic displacements $d_{x}(r)$, $d_{y}(r)$, (b) local
strains $u_{xx}(r)$, $u_{yy}(r)$, (c) $u_{xy}(r)$ and local polarization $P_{x}(r)$ for the
alternating domain-stripe solutions of the model inhomogeneous ferroelectric
in 2-dimension.

Figure 8. Contourplot of the inelastic scattering intensity for
the alternating domain-stripe solution discussed in Section 6.

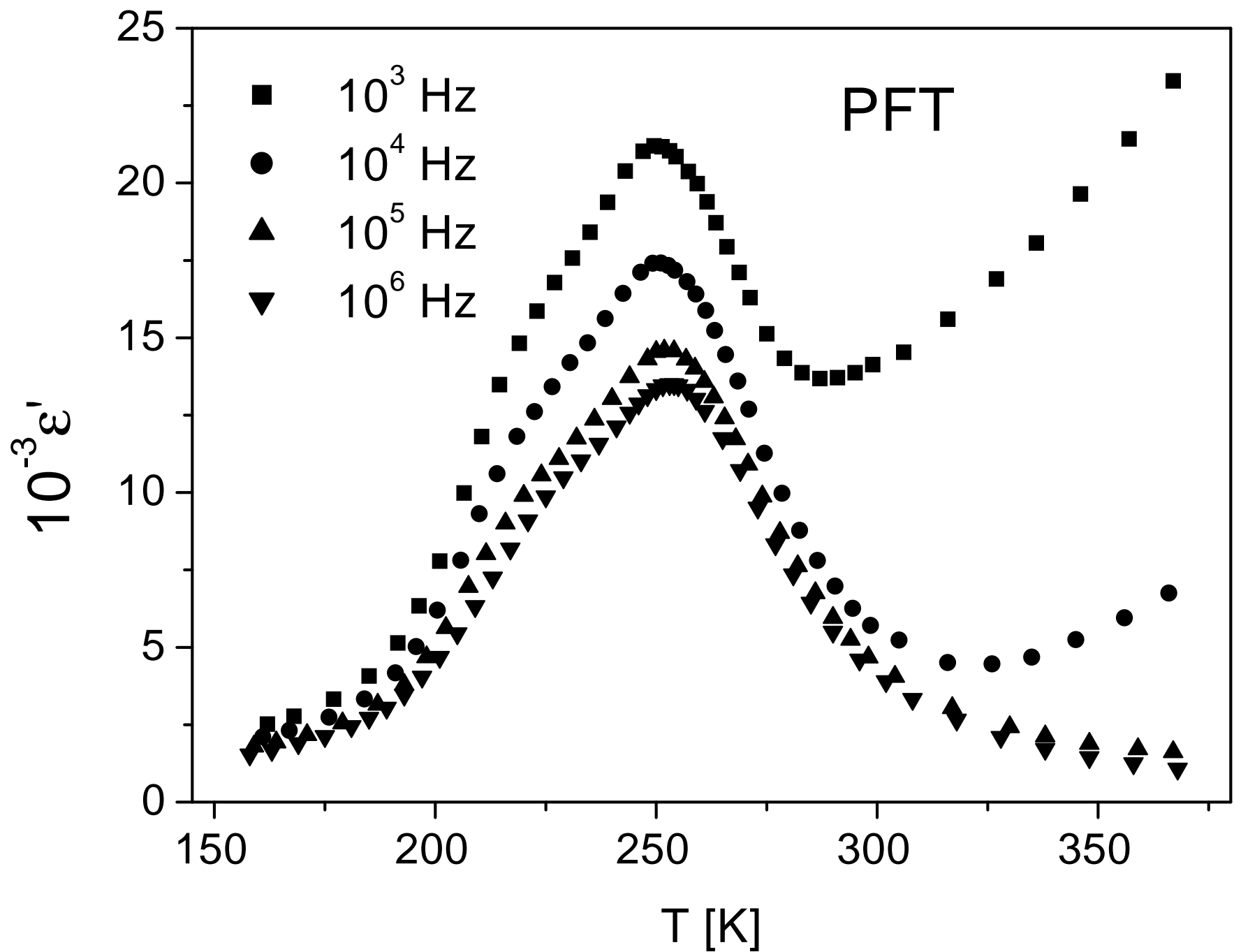
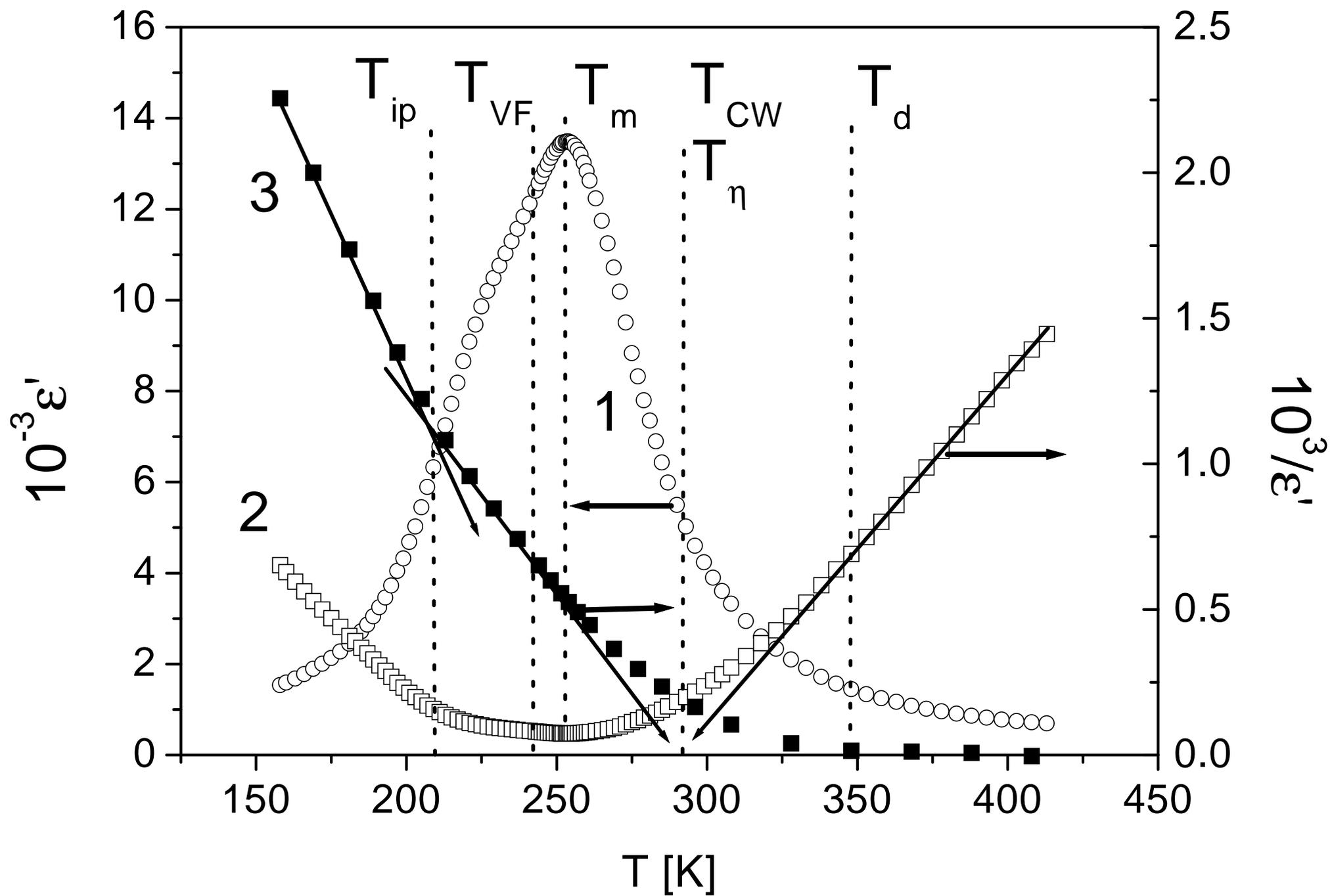
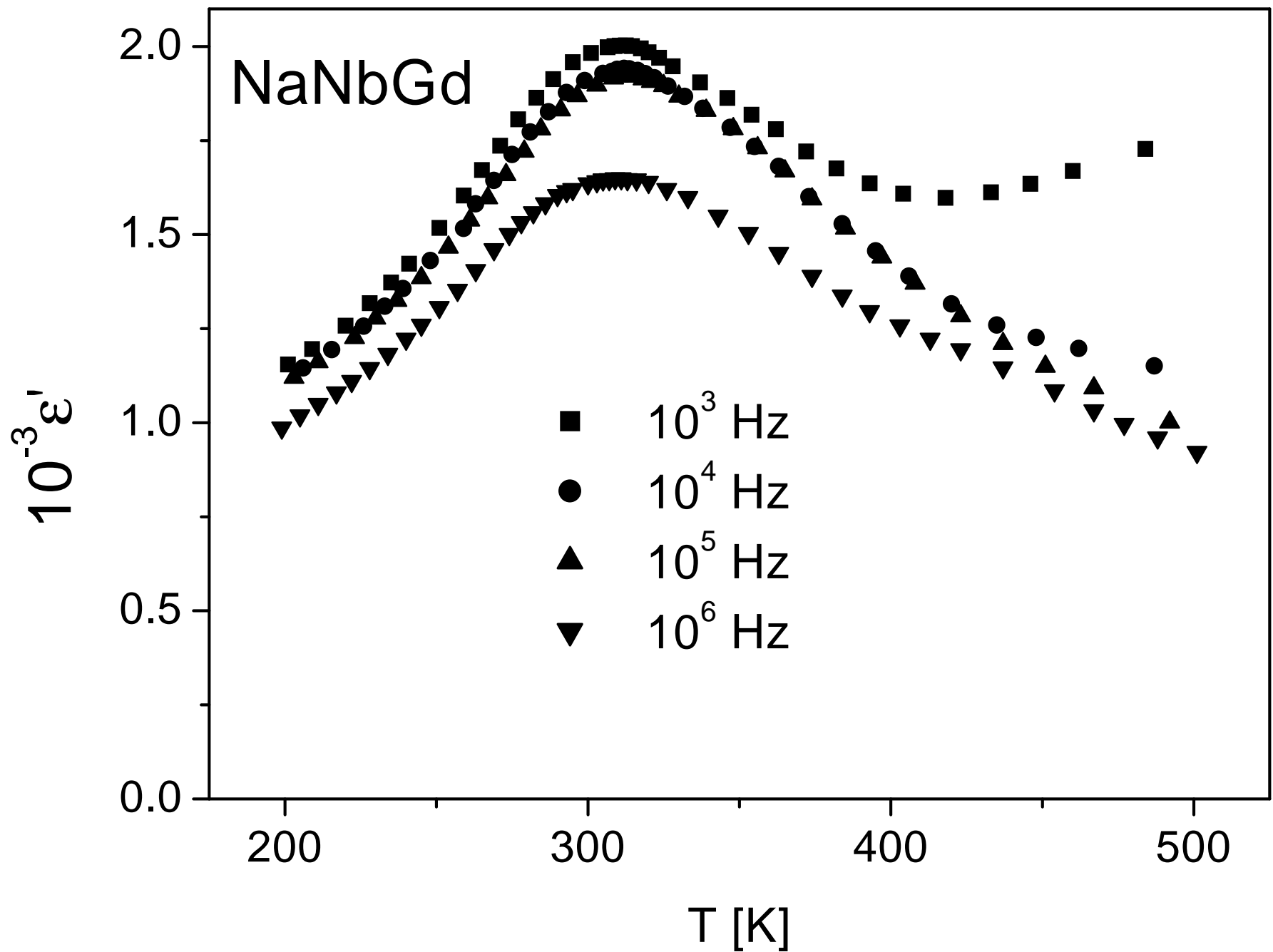
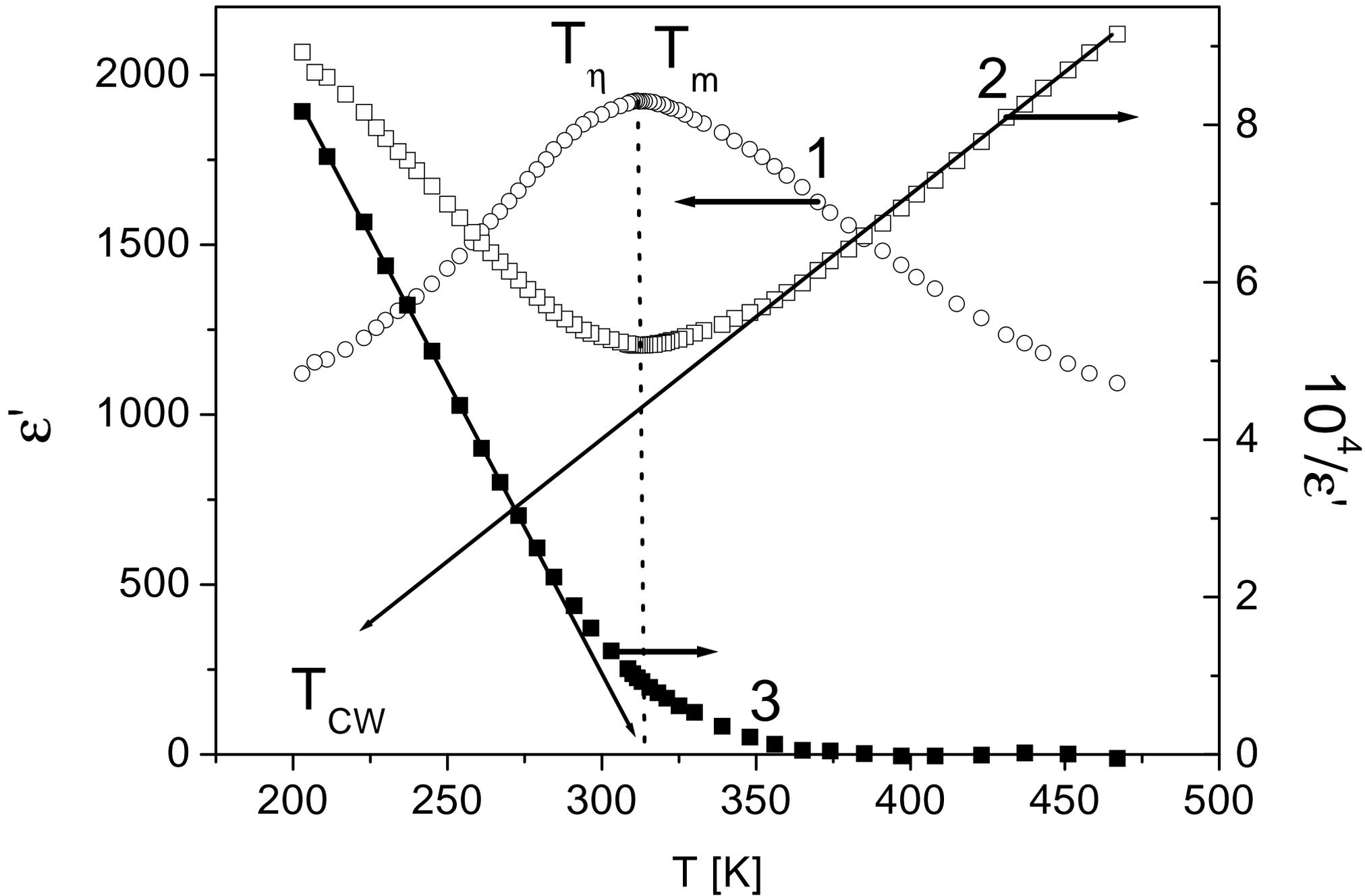
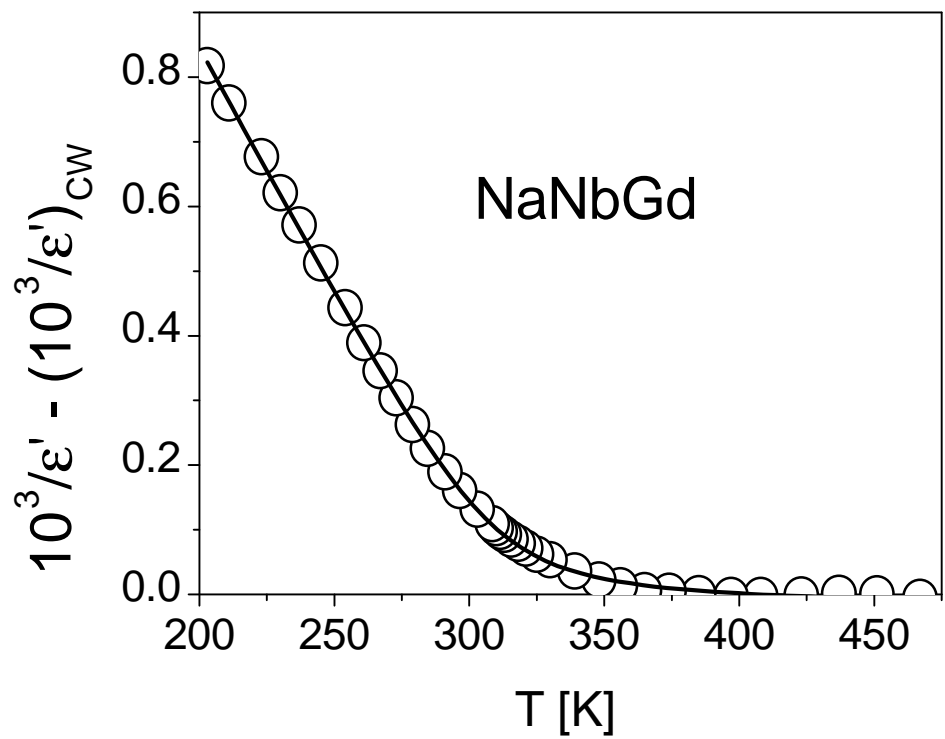
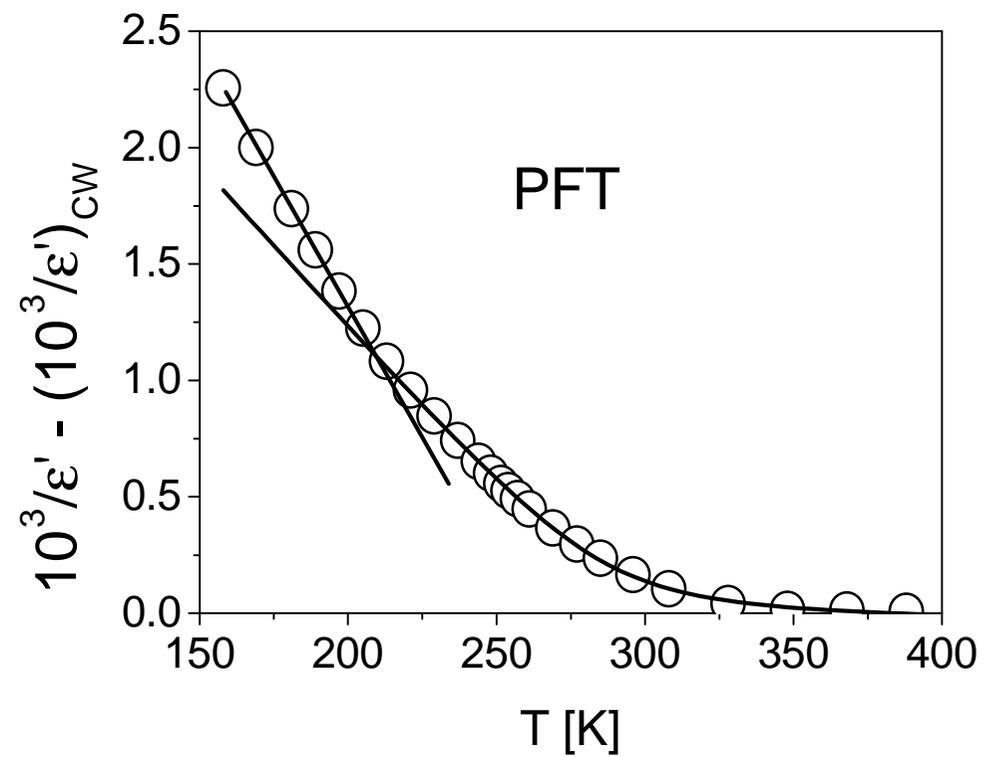
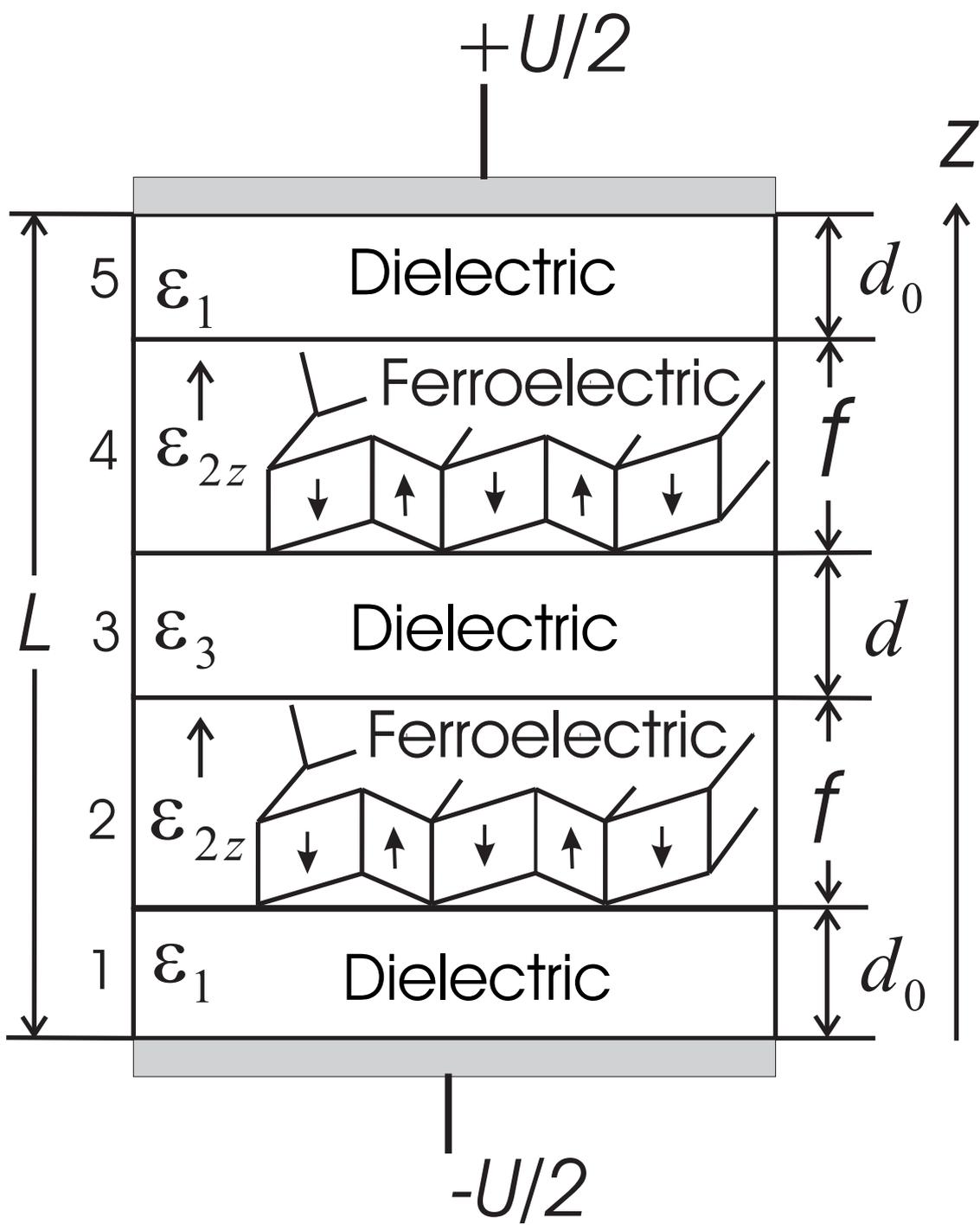
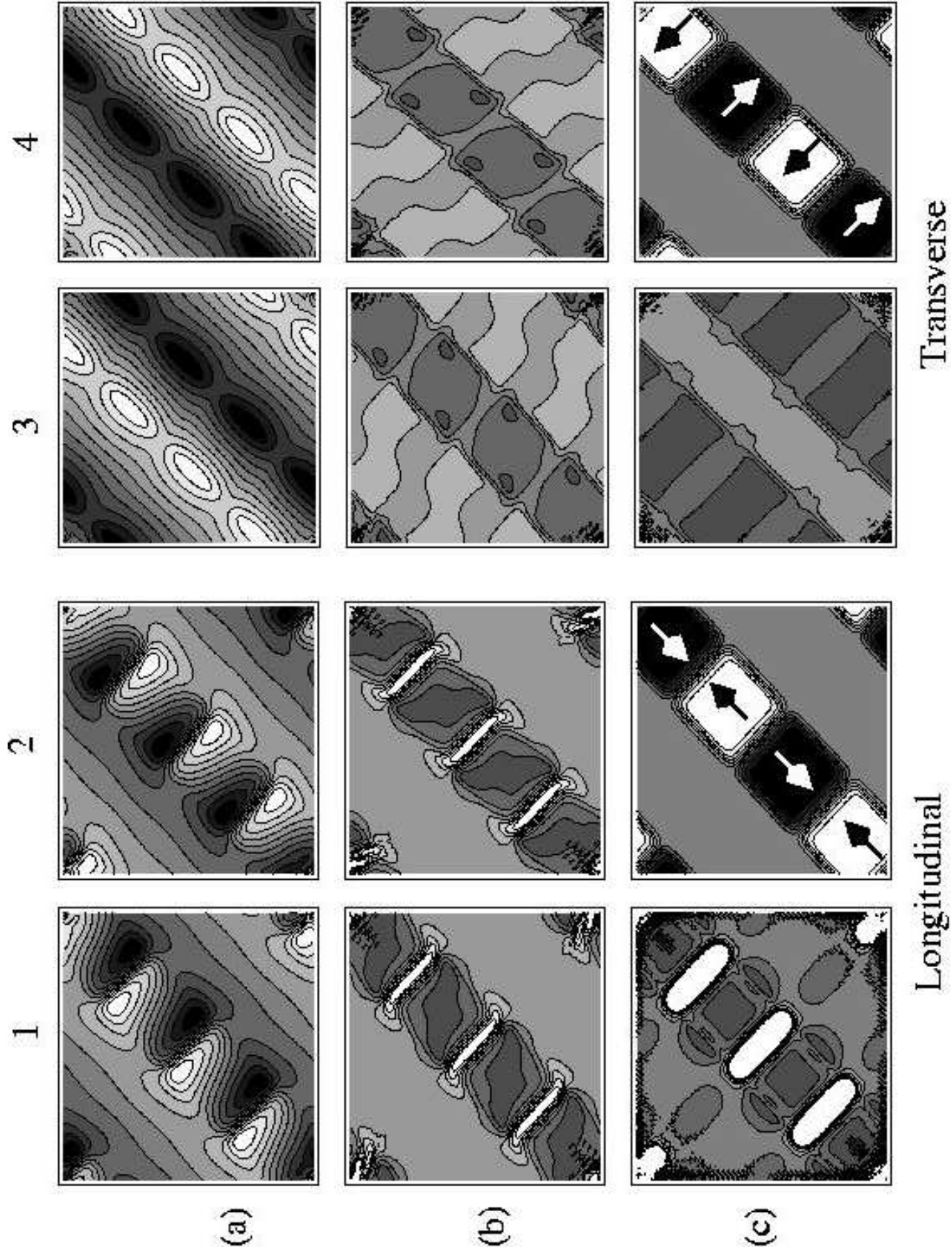
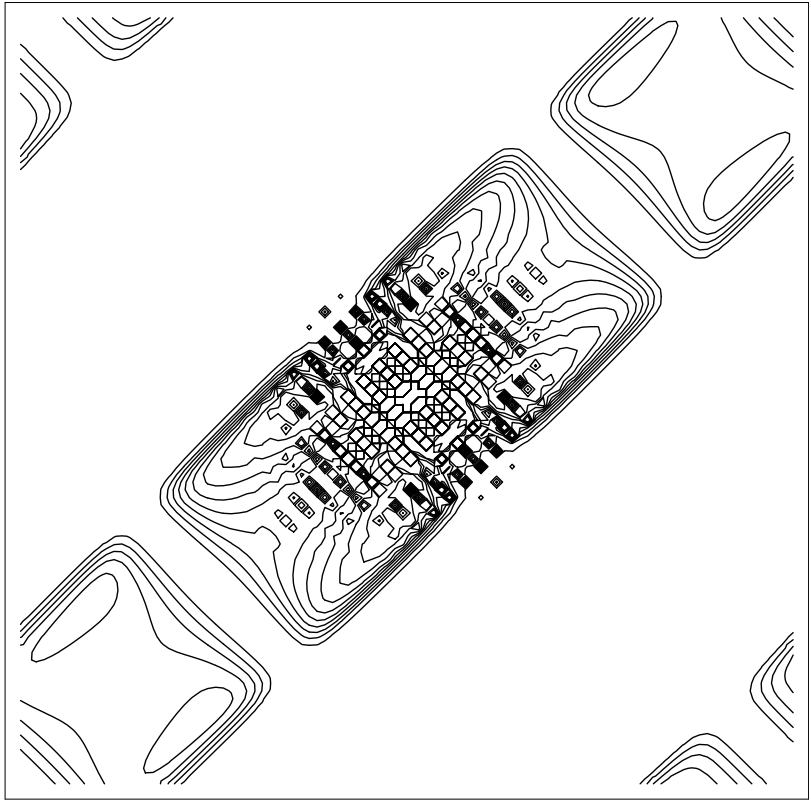
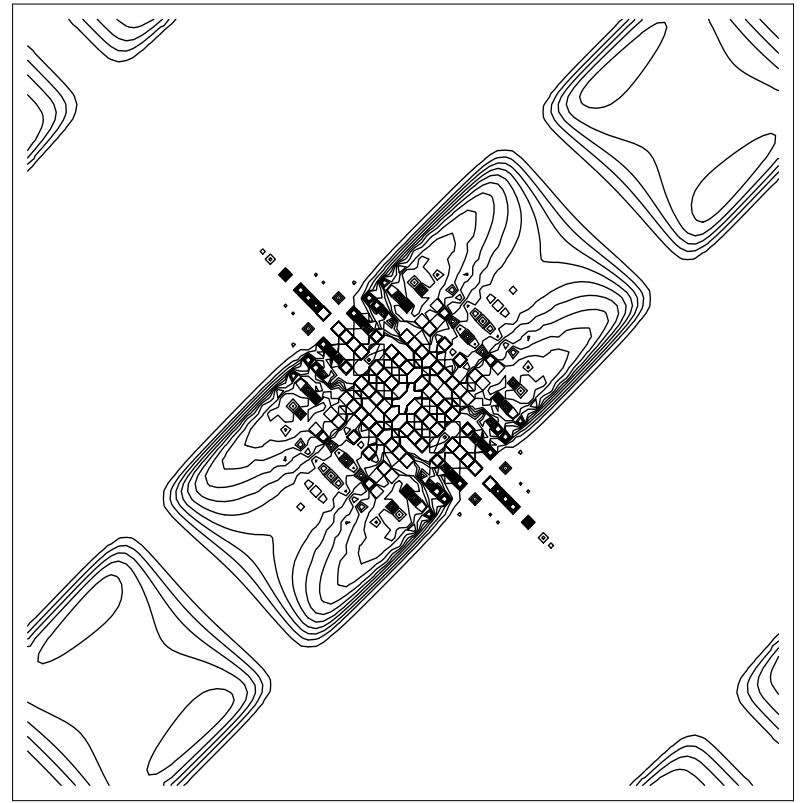
\end{document}